# Kinetic limitation of chemical ordering in $Bi_2Te_{3-x}Se_x$ layers grown by molecular beam epitaxy

S. Schreyeck[1], K. Brunner[1], A. Kirchner[1], U. Bass[1], S. Grauer[1], C. Schumacher[1], C.Gould[1], G. Karczewski[2], J. Geurts[1], and L. W. Molenkamp[1]

[1] Physikalisches Institut, Experimentelle Physik 3 and Wilhelm-Conrad Röntgen-Research Center for Complex Material Systems, Universität Würzburg, Am Hubland, D-97074 Würzburg, Germany
[2] Institute of Physics, Polish Academy of Sciences, Al. Lotników 32/46, 02-668 Warsaw, Poland

E-mail: sschreyeck@physik.uni-wuerzburg.de, brunner@physik.uni-wuerzburg.de

We study the chemical ordering in $Bi_2Te_{3-x}Se_x$ grown by molecular beam epitaxy on Si substrates. We produce films in the full composition range from $x = 0$ to 3, and determine their material properties using energy dispersive X-ray spectroscopy, X-ray diffraction and Raman spectroscopy. By fitting the parameters of a kinetic growth model to these results, we obtain a consistent description of growth at a microscopic level. Our main finding is that despite the incorporation of Se in the central layer being much more probable than that of Te, the formation of a fully ordered *Te-Bi-Se-Bi-Te* layer is prevented by kinetic of the growth process. Indeed, the Se concentration in the central layer of $Bi_2Te_2Se_1$ reaches a maximum of only ≈ 75% even under ideal growth conditions. A second finding of our work is that the intensity ratio of the *0 0 12* and *0 0 6* X-ray reflections serves as an experimentally accessible quantitative measure of the degree of ordering in these films.

# 1. Introduction

$Bi_2Te_{3-x}Se_x$ alloys, which continue to be of great interest in the field of thermoelectric devices, [1] have also more recently been shown to be topological insulators hosting a topologically protected spin polarized surface state.[2,3] For use as a topological insulator (TI), the mixed alloy with $x \approx 1$ combines positive features of each of the binary materials $Bi_2Te_3$ and $Bi_2Se_3$. Like $Bi_2Se_3$, it offers a relatively large band gap of $\approx 0.3$ eV with the Dirac point of the TI surface state at an energy within the bulk band gap.[4,5] At the same time it shares $Bi_2Te_3$'s property of being relatively insensitive to vacancy defects, allowing the material to be grown with lower bulk carrier density than $Bi_2Se_3$.[4, 6]

The ideal ternary compound $Bi_2Te_2Se$ has an ordered tetradymite-like structure consisting of quintuple layers (QL) *Te-Bi-Se-Bi-Te* (see inset of Figure 1). Atoms within the QL are chemically bonded to each other, whereas the QLs are bonded to each other through weaker van-der-Waals forces.

When considering growth of a real crystal it is convenient to describe the structure of the QL as *VI(1)-Bi-VI(2)-Bi-VI(1)*, where both Se and Te can occupy either the *VI(1)* or *VI(2)* sites. The ordering is then driven by the large electronegativity of Se compared to Te which greatly favors its incorporation into the *VI(2)* sites where it forms six chemical bonds to Bi atoms compared to the only three chemical bonds of the *VI(1)* sites.[7] As this selectivity mechanism is not perfect, a method to determine the degree of ordering of a film becomes an important tool for the study of these materials. This is experimentally relevant since structural ordering maximizes the band gap while disorder enhances alloy scattering of electrons and phonons.[1,6]

In this paper we study $Bi_2Te_{3-x}Se_x$ layers grown by molecular beam epitaxy (MBE) on Si(111) substrates with $x$ values ranging from 0 to 3. By combining structural and compositional analysis and a kinetic growth model, we show how the degree of ordering can be determined, and that for the case of $x = 1$, it is kinetically limited to a value of $\approx 75\%$.

## 2. Experimental Section

Molecular beam epitaxy of epitaxial $Bi_2Te_{3-x}Se_x$ layers is performed by co-deposition of elemental materials (6N purity) under ultra-high vacuum (UHV) conditions (base pressure < $10^{-10}$ mbar) on H-passivated Si(111) substrates at a substrate temperature of 300°C. All fluxes $f \propto BEP \cdot \sqrt{T/M} / \eta$ are determined from the source temperature $T$, the beam equivalent pressure (*BEP*), the Bayard-Alpert gauge sensitivity $\eta$ and the molecular mass $M$ by assuming fluxes of tetramer molecules.[8,9] A layer growth rate $r$ of 1.0 QL per minute is deduced from the layer thicknesses of about 70 nm (determined by profilometry on a mesa structure) with a 70 minutes growth time. Given that all growths are under group *VI* rich conditions, the growth rate $r$ is limited by the Bi flux. The absolute flux of Bi can be determined using the constant layer growth rate of 1 QL (which contains 2 Bi monolayers (ML)) per minute. In our studies, the Bi and Te fluxes are kept constant ($f_{Bi}$ = 2 ML/min and $f_{Te}$ = 84 ML/min) while the Se flux $f_{Se}$ is varied from 0 to 250 ML/min to obtain a series of samples covering the full composition range. Additionally a pure $Bi_2Se_3$ layer is deposited without Te flux. The Se content $x$ of the layers is measured to an accuracy of about ±0.05 by energy dispersive X-ray spectroscopy (EDX) of the Bi, Te and Se emission lines. High resolution X-ray diffraction (XRD) is performed with a Panalytical X`Pert diffractometer equipped with a Cu-K$_{\alpha 1}$ source. Raman spectroscopy is performed at room temperature with a low power laser with wavelength $\lambda$ = 633 nm in order to avoid heating and degradation of the layers.

## 3. Results and Discussion
### 3.1. Chemical Composition and Kinetic Growth Model

The Se/Te content ratio $x/(3-x)$ of the $Bi_2Te_{3-x}Se_x$ layers, as measured by EDX, is given as the data points in Figure 1 as a function of the Se/Te flux ratio $f_{Se}/f_{Te}$. The Se/Te content ratio is up to some 8 times higher than the flux ratio. This suggests that the incorporation of Se is much more probable than that of Te. Such a behavior is also known for MBE of other alloy systems such as zinc-blende $ZnTe_{1-x}Se_x$.[10,11] It can be explained by the high electronegativity of Se compared to Te. This behavior is consistent with results from bulk crystal rods grown by the Bridgman-Stockbarger method from a stoichiometric $Bi_2Te_2Se_1$ melt. There, strong longitudinal gradients in Se content also indicate a preferential incorporation of Se.[12] The nonlinear increase of the content ratio as a function of flux ratio in Figure 1 shows three distinct regions. A nearly linear increase of Se/Te content ratio at low flux ratios (see left-hand inset of Figure 1) is followed by a region of increased slope for a flux ratio of $f_{Se}/f_{Te} > 0.1$ and a region with decreasing slope at high flux ratios.

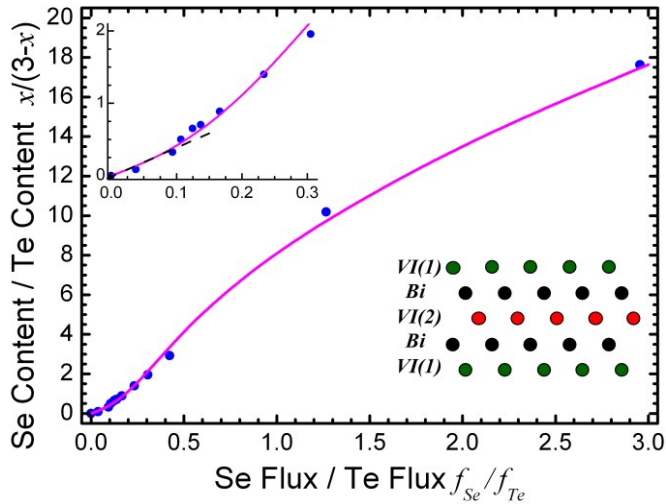

**Figure 1:** Se/Te content ratio $x/(3-x)$ (blue dots) measured by EDX for MBE grown 70 nm thick $Bi_2Te_{3-x}Se_x$ layers on Si(111) as a function of the flux ratio $f_{Se}/f_{Te}$. The Se/Te content ratio calculated with a kinetic model description of adsorption, desorption and incorporation of Se and Te at non-equivalent group *VI* sublayer sites *VI(1)* and *VI(2)* in the QL structure (see right-hand inset) according to Equation 1 is given by the solid curve. A close-up of the region with low $f_{Se}/f_{Te}$ flux ratios and a linear approximation are shown in the left-hand inset.

In order to gain deeper insight into the MBE growth, and based on established models describing growth of *II-VI* and *III-V* zinc-blende materials[13, 14], we developed a kinetic model that describes the details of incorporation of adsorbed Se and Te, which each compete for two non-equivalent types of sublayer sites: one *VI(2)* site and two *VI(1)* sites (as identified in the schematic of Figure 1).

Our kinetic model is schematically described in Figure 2a and considers the adsorption and desorption of Se, Te, and Bi molecules as well as the preferential incorporation of Se at non-equivalent sites *VI(1)* and *VI(2)* in the stationary state. All Bi supplied is assumed to be adsorbed at the surface and nearly instantly incorporated at the Bi lattice sites with a rate $r_{Bi} = f_{Bi} = 2$ ML/min, implying the Bi surface coverage is negligibly small and can be set to $n_{Bi} = 0$. The four free parameters in the model are the desorption coefficients $d_{Se}$ and $d_{Te}$ for Se and Te, respectively, and the ratio $S_i$ of the incorporation probability for adsorbed Se compared to Te on site *VI(i)*.

The resulting stationary solutions are given in Equation 1 and are described as follows. Equation 1a (1b) describes the equilibrium between the surface adsorption rate (left-hand side) and the sum of incorporation and desorption rate (right-hand side) for Se (Te). The growth rates of the central sublayer, site *VI(2)*, and the two edge sublayers, sites *VI(1)*, are also fixed at $r_2 = f_{Bi}/2$ and $r_1 = f_{Bi}$, respectively, due to the QL structure. The formation of point defects such as $Se_{Bi}$ and $Te_{Bi}$ antisites and vacancies as well as decomposition of the layer play a minor role and are not considered.

As the total surface coverage (in ML) by Se $n_{Se}$ and Te $n_{Te}$ is limited to 1, i.e. $n_{Se} + n_{Te} \leq 1$, adsorption of Se and Te from the supplied fluxes results in the left-hand terms in Equation 1a and 1b, respectively. The squaring of the in-parenthesis surface coverage term comes from the fact that group *VI* materials are known to evaporate from effusion cells as molecules and adsorption requires at least two neighboring free surface sites.[13, 14] Similarly, desorption of Se or Te as molecules requires the occupation of two neighboring surface sites and contributes the square in the right-hand terms in Equation 1a and 1b. The incorporation probability of an adsorbed Se atom into a sublayer site $VI(i)$ ($i$ = 1, 2) exceeds that of an adsorbed Te atom by a factor $S_i$. The Se content $x_1$ ($x_2$) at sublayer sites $VI(1)$ (site $VI(2)$) is equal to the incorporation efficiencies $E_{Se(1)}$ ($E_{Se(2)}$) defined in Equation 1c and 1d. They depend on the factors $S_i$ and the surface coverages that limit the supply of Se and Te. For MBE growth of a random alloy crystal structure with just one type of site $VI(i)$, Equation 1 would result in a Se/Te content ratio proportional to $S_i$ and the ratio of surface coverages $n_{Se}/n_{Te}$.

For a $Bi_2Te_{3-x}Se_x$ crystal with the non-equivalent sites $VI(1)$ and $VI(2)$, the model results in the two rate equations 1a and 1b. These are nonlinearly coupled through the surface coverages $n_{Se}$ and $n_{Te}$ that depend on the fluxes $f_{Se}$ and $f_{Te}$. The desorption coefficients $d_{Se}$ and $d_{Te}$ influence the surface coverages and consequently the Se content $x$ at low as well as at high flux ratios $f_{Se}/f_{Te}$. The incorporation of Se at site $VI(2)$ compared to sites $VI(1)$ is characterized by the site selectivity of Se $s = S_2/S_1$.

$$(1-n_{Se}-n_{Te})^2 f_{Se} = E_{Se(1)} f_{Bi} + E_{Se(2)} f_{Bi}/2 + d_{Se} n_{Se}^2 \tag{1a}$$

$$(1-n_{Se}-n_{Te})^2 f_{Te} = E_{Te(1)} f_{Bi} + E_{Te(2)} f_{Bi}/2 + d_{Te} n_{Te}^2 \tag{1b}$$

$$E_{Se(i)} \equiv S_i n_{Se}/(n_{Te}+S_i n_{Se}) = x_i \quad (i=1, 2) \tag{1c}$$

$$E_{Te(i)} \equiv 1 - E_{Se(i)} \tag{1d}$$

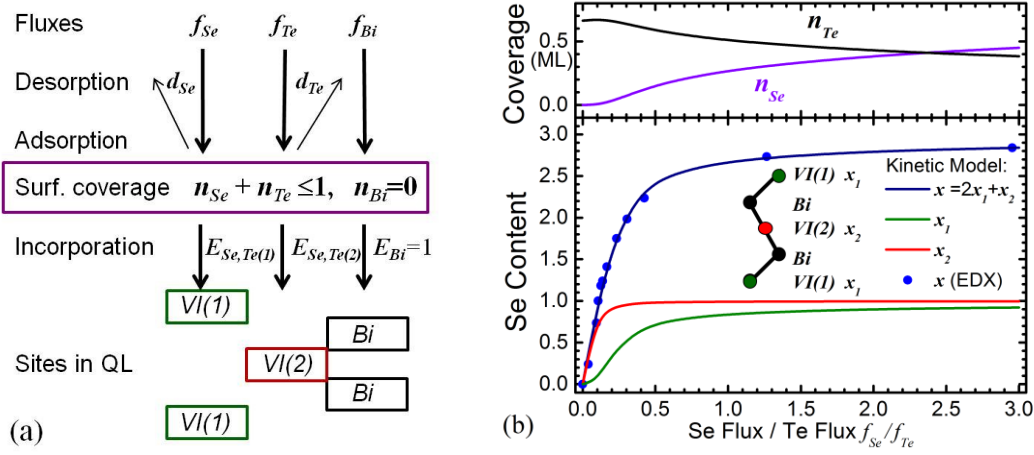

**Figure 2:** (a) Schematics of the kinetic model (see Equation 1) for MBE growth of $Bi_2Te_{3-x}Se_x$ with QL structure *VI(1)-Bi-VI(2)-Bi-VI(1)*. Adsorption, desorption of molecules of Se, Te, Bi and preferential incorporation of Se at non-equivalent sites *VI(1)* and *VI(2)* are considered in the stationary state. (b) Lower part: Calculated Se contents $x_1$, $x_2$ and $x$ of the sublayer sites *VI(1)* (green), the center sublayer site *VI(2)* (red) and the total QL (blue), depending on Se/Te flux ratio. The values of Se content $x$ measured by EDX are given by blue dots. Upper part: The calculated surface coverages by Se $n_{Se}$ and by Te $n_{Te}$ versus Se/Te flux ratio.

In order to fit our model to the experimental results, we vary the four free model parameters $S_1$, $S_2$, $d_{Se}$, $d_{Te}$ in Equation 1. The Se/Te content ratio, which results from numerically solving the coupled Equation 1 (such as $n_{Se}$ and $n_{Te}$), in dependence on Se/Te flux ratio is plotted in Figure 1 for a given set of parameters ($S_1$ = 10, $S_2$ = 220, $d_{Se}$ = 21 ML/min and $d_{Te}$ = 15 ML/min) and agrees well with the EDX results. This suggests that the main kinetic aspects for MBE growth of $Bi_2Te_{3-x}Se_x$ layers are captured by the model. These parameters imply a much higher probability of incorporation of Se compared to that of Te, especially at site *VI(2)*, resulting in a site selectivity of Se $s$ = 22.

It should be noted that the set of parameters used above is not unique to fitting the EDX data, but was chosen as follows: The desorption coefficient $d_{Te}$ = 15 ML/min was determined from the slope of the curve at low flux ratio $f_{Se}/f_{Te}$ < 0.1 (see left-hand inset of Figure 1), at which, according to Equation 1, adsorption and consequently incorporation of Se is limited by the surface coverage with Te ($n_{Te}$ = 0.66). A desorption coefficient of Se $d_{Se}$ = 21 ML/min > $d_{Te}$ was deduced from the behavior at high $f_{Se}/f_{Te}$ and describes a realistic scenario, where the vapor pressure of Se is higher than that of Te. The $S_i$ values are consistent with Boltzmann factors calculated from chalcogen antisite defect formation energies (Te$_{Se}$ and Se$_{Te}$) in ideally ordered *Te-Bi-Se-Bi-Te*.[15] The chosen model parameters, especially the site selectivity $s$ = 22, are supported by the consistency of the results of the kinetic growth model with the experimental XRD and Raman results that are described in subsections 3.2. and 3.3..

In addition to describing the total Se content $x$ described in Figure 1, our model also yields information about the Se content on each of the sublayers. In Figure 2b we plot the calculated Se content in sublayer sites *VI(1)* $x_1$ and sublayer site *VI(2)* $x_2$ together with the total amount of Se in a QL $x = 2x_1 + x_2$ versus the flux ratio $f_{Se}/f_{Te}$. The data from Figure 1 is replotted here for comparison. The calculated Se content $x_2$ of sublayer *VI(2)* quickly increases with increasing Se flux and starts to saturate at about $f_{Se}/f_{Te}$ = 0.1.

Equipped with this information, the non-linear behavior seen in Figure 1 can now be understood. The calculated Se surface coverage $n_{Se}$ plotted in Figure 2b is small for low $f_{Se}/f_{Te}$ < 0.1, due to the efficient Se incorporation at site *VI(2)*, but $n_{Se}$ becomes significant as $x_2$ starts to saturate. Consequently, $n_{Te}$ decreases (see Figure 2b) and the incorporation of Se at site *VI(1)* (and $x_1$) starts to increase strongly for $f_{Se}/f_{Te}$ > 0.1. In plain terms, this means that for low fluxes, the Se incorporates (nearly) only in the central layer, and once the flux ratio reaches ≈ 0.1 it gains access to the two outer sublayers.

An important element of our model is that it allows us to quantify the degree of order in the films, which we define as the fraction of Se at the central sublayer site *VI(2)* $x_2/x$. It has a maximum value of $x_2/x = s/(s+2) = 92\%$ at very low $f_{Se}/f_{Te}$ (and thus very low total Se content $x$) but as $f_{Se}/f_{Te}$ increases the incorporation efficiency $E_{Se(2)} = x_2$ starts to saturate. Thus the incorporation of Se at sites *VI(1)* becomes significant and precludes reaching an ideally ordered Bi$_2$Te$_2$Se$_1$ layer. Indeed, MBE-grown Bi$_2$Te$_2$Se$_1$ layers are limited to a degree of chemical order $x_2/x = 75\%$ despite the high site selectivity of Se $s = 22$. This mechanism limiting the chemical order appears to be general in the sense that the degree of order is determined only by the site selectivity $s$. The analytical solution of Equation 1c *(i = 1, 2)* with the condition $2x_1 + x_2 = 1$ reveals that the degree of order in Bi$_2$Te$_2$Se$_1$ expressed by $x_2$ increases slowly with site selectivity $s$ and is approximately given by $x_2 \approx \left(1 - \sqrt{2/s}\right)$ for very large $s$ ($S_2 > 100 S_1$). As a main result, the sublayer compositions $x_1$ and $x_2$ for any total Se content $x$ are determined only by the site selectivity $s$. A variation of substrate temperature is expected to change the values of $S_i$ and hence also $s = S_2/S_1$, but has minor influence on the degree of order. Variations in other model or experimental parameters (retaining $f_{Se} + f_{Te} \gg f_{Bi}$) affect the surface coverages and the Se content $x$ in dependence on the fluxes, but do not influence the degree of order at any given Se content, e.g. $x = 1$.

**3.2. Chemical Order Analysis by XRD**

A direct measurement of the Se content of the sublayers $x_1$ and $x_2$ is extremely difficult. Significant information can however be gained from analyzing its influence on XRD measurements. Figure 3 shows θ-2θ scans of layers of various compositions. Several symmetric *0 0 l* XRD reflections are observed confirming that all layers grow with the c-axis parallel to the surface normal of the Si(111) substrate and have single phase tetradymite-like structure.

The *0 0 12* peak varies in intensity depending on the Se content *x* whereas the *0 0 6* peak is nearly constant in intensity. The peaks shift to larger diffraction angles with increasing Se content. The out of plane lattice parameter c of the hexagonal unit cell calculated from the peak position is plotted as the data points in Figure 4 against Se content. For x = 0, 1 and 3, the lattice parameter *c*, which corresponds to the height of 3 QLs, is consistent with literature values for $Bi_2Te_3$, ordered $Bi_2Te_2Se_1$ and $Bi_2Se_3$, respectively.[16, 17] These values of *c* and those of the in-plane lattice parameter *a* determined from asymmetric reflections (not shown) confirm that the layers are relaxed and that lattice strain can be neglected. The dependence of the lattice parameter *c* on Se content clearly deviates from linearity (Vegard`s law) and reveals a bowing for *x* > 1, as also observed for bulk crystals.[18] This bowing is assigned to the non-linear change in *VI(1)-VI(1)* separation $c_{vdW}$ between the QL as a function of increasing $x_1$. The van-der-Waals bonds of mixed atomic pairs Se-Te are weaker than those of Te-Te or Se-Se pairs. The deviation from Vegard`s law is thus described by a term proportional to the probability $x_1(1 - x_1)$ of van-der-Waals bonded Se-Te pairs and is maximal for $x_1 = 0.5$.[18, 19]

The lattice parameter *c* of $Bi_2Te_{3-x}Se_x$ for arbitrary sublayer compositions can be interpolated between the known literature values for $Bi_2Te_3$, ordered $Bi_2Te_2Se_1$ and $Bi_2Se_3$, using:

$c(x_1, x_2) = 30.42 \text{ Å} - x_2\, 0.56 \text{ Å} - x_1\, 1.24 \text{ Å} + x_1(1-x_1)\, 0.6 \text{ Å}$. (2)

The red curve in Figure 4 is the calculated value of *c* using the sublayer compositions $x_1$, and $x_2$ from our kinetic growth model. Its agreement with experimental values is well within experimental accuracy. A comparison of the red curve with the blue curve, which is calculated for perfect order ($x_2 = x$, $x_1 = 0$ for $x \leq 1$; $x_2 = 1$, $x_1 = (x - 1)/2$ for $x > 1$), and with the black curve describing a random alloy ($x_1 = x_2 = x/3$), shows that all three curves are similar and partially overlap. Obviously, the lattice parameter *c* is rather insensitive to the degree of order and these results cannot reliably discriminate between the different degrees of ordering.

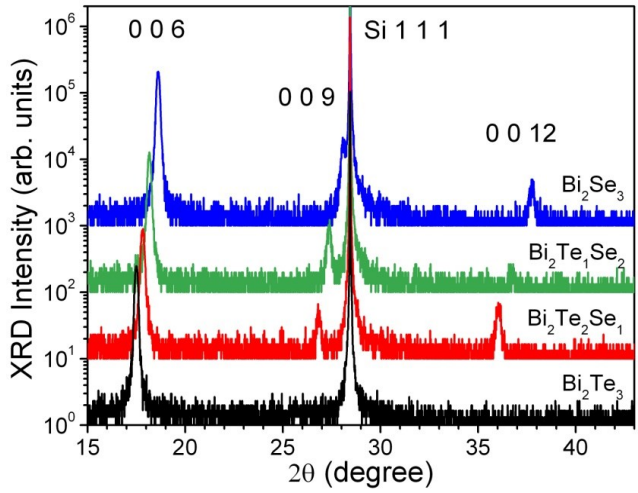

**Figure 3:** Wide angular range XRD θ-2θ scans from various 70 nm $Bi_2Te_{3-x}Se_x$/Si(111) layer structures. The indices *0 0 l* of $Bi_2Te_{3-x}Se_x$ reflections and the *1 1 1* Si substrate reflection are indicated. The diffractograms are vertically shifted for clarity.

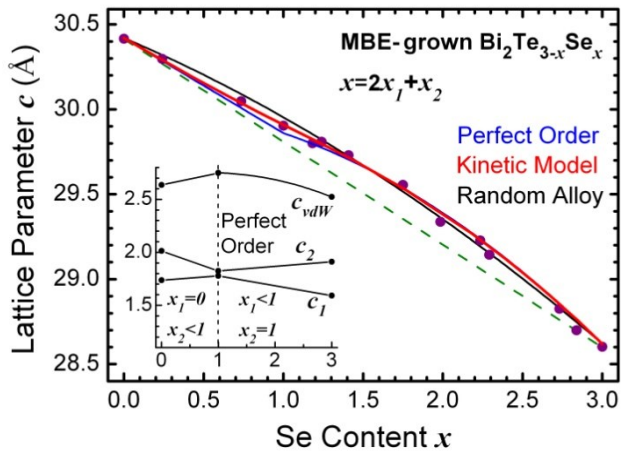

**Figure 4:** Lattice parameter *c* determined from the θ-2θ scans of the *0 0 l* reflections of the $Bi_2Te_{3-x}Se_x$ layers (violet dots) versus Se content *x*. The curves are plotted using Vegard`s law (dashed green line) or Equation 2 for layers with perfect order (blue curve), a random alloy (black curve) and the kinetic model (red curve). The inset shows the sublayer separations in perfectly ordered $Bi_2Te_{3-x}Se_x$ versus *x*.

The separations $c_1$, $c_2$ and $c_{vdW}$ between neighboring sublayers *VI(1)-Bi*, *Bi-VI(2)* and van-der-Waals bonded sublayers *VI(1)-VI(1)*, respectively, can be calculated from the lattice parameter $c = 3(2c_1 + 2c_2 + c_{vdW})$ and the Wyckoff positions of the atoms in $Bi_2Te_3$, ordered $Bi_2Te_2Se_1$ and $Bi_2Se_3$.[16, 17] These sublayer separations are plotted in the inset of Figure 4 assuming perfect order and they reveal that all the sublayer separations depend on Se content as well as on the degree of ordering, with each of the dependences being non-monotonic.

Se replacing Te at sublayer site $VI(2)$ reduces the separation $c_2$ to the neighboring Bi sublayers due to its smaller covalent radius and its higher electronegativity (2.4) than Te (2.1). This increase in ionic character of the $VI(2)$-Bi bond transfers electron charge into this bond from the $VI(1)$-Bi bond.[7] The resulting reduction of polarity of the $VI(1)$-Bi bond increases the separation $c_1$ (and $c_{vdW}$) with increasing $x_2$. Conversely, replacing Te by Se at site $VI(1)$ causes a decrease of $c_1$ and $c_{vdW}$ (with bowing) but an increase of $c_2$.

We now consider the influence of the ordering on the intensity of the XRD reflections. The data points in Figure 5 show the integrated intensity of the *0 0 12* reflection as a function of Se content. These have been normalized to the intensity of the *0 0 6* reflection, which is nearly independent of Se content, in order to reduce experimental inaccuracies. The intensity ratio $I_{0\,0\,12}/I_{0\,0\,6}$ shows an oscillatory behavior with a pronounced maximum at $x \approx 1$ and a smaller one at $x = 3$ (i. e. $Bi_2Se_3$). Minima in intensity occur for $x = 0$ and at $x \approx 2.5$ where the *0 0 12* peak intensities are comparable to background signal. A similar behavior was observed for $Bi_2Te_{3-x}Se_x$ layers grown by metalorganic vapor phase epitaxy.[20]

We have calculated the structure factors $S_{0\,0\,12}$ and $S_{0\,0\,6}$ for these reflections using the layer separations discussed above and atomic form factors of the sublayers $VI(i)$ determined by linear interpolation between those of Te and Se based on the composition $x_i$ calculated from our kinetic model.[21] The ratio of squared structure factors, which is proportional to the normalized intensity, is plotted as the red curve in Figure 5. The experimental data are in good agreement with the calculated curve (to within a scaling factor[22] which unites the right and left axis in the figure and which was chosen to make the calculation for pure $Bi_2Se_3$ fit to the data) for partial ordering determined by the site selectivity $s = 22$ of the kinetic model.

The calculated intensity ratios for $Bi_2Te_{3-x}Se_x$ layers with perfect order (blue curve) and for a random alloy (black curve) are also shown for comparison. For $Bi_2Te_2Se_1$, perfect order gives a nearly 3-times higher intensity ratio than is experimentally observed, while the intensity ratio is small and monotonic in $x$ for the random alloy.

The inset of Figure 5 again shows the ratio of squared structure factors, now calculated for $Bi_2Te_2Se_1$ ($x = 1$) layers with varying degree of order $x_2$. The intensity ratio is small for a random or nearly random alloy, and starts to increase significantly and monotonically at $x_2 \approx 0.5$. The red dot marking the intensity ratio from our kinetic model corresponds to an ordering $x_2$ of 0.75, where permitting an uncertainty of ±1% in the literature values of sublayer separations $c_1$ and $c_2$ (at constant $c$) can shift the determined degree of order by up to ±0.05. Thus, the XRD measurement of the intensity ratio $I_{0\,0\,12}/I_{0\,0\,6}$ of a $Bi_2Te_2Se_1$ layer (compared to that of a $Bi_2Se_3$ reference layer for instrumental calibration) is a tool for analyzing the degree of structural order.

It is worth noting that due to the large slope in the inset of Figure 5, assuming an uncertainty of ±20% on the measured peak intensity ratio still yields a result of $x_2 = 0.75 \pm 0.04$. Even considering statistical fluctuation for the relatively weak *0 0 12* peak and difficulties in perfectly applying XRD correction factors that vary slightly with Se content $x$ due to the shifts in Bragg angles, a ±20% uncertainty estimate is extremely conservative. This highlights the sensitivity of this method for characterizing the degree of ordering in at least relatively highly ordered layers.

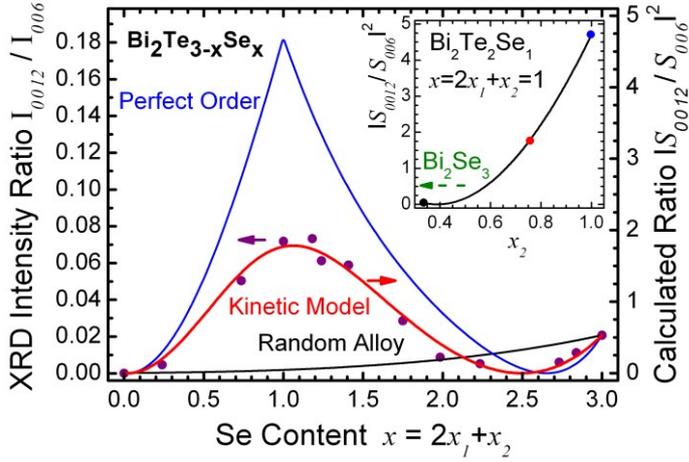

**Figure 5:** Ratio of the integrated *0 0 12* and *0 0 6* intensities measured by XRD (violet circles) and corresponding squared ratio of calculated structure factors versus Se content for the random alloy (black), full order (blue), and partial order according to the kinetic model (red curve). The inset shows the ratio of squared structure factors for $Bi_2Te_2Se_1$ ($x = 1$) depending on Se content of the center sublayer $x_2$, and the same ratio for $Bi_2Se_3$ (dashed green arrow) for reference. The data points for the three degrees of order considered in the main figure are marked by circles of corresponding color.

### 3.3. Impact of Chemical Order on Lattice Dynamics

Further confirmation of our interpretation of chemical ordering is provided by the set of Raman spectra in Figure 6a. The spectra are observed in backscattering with parallel polarizations of the exciting and the scattered light. They show four optical phonon modes: the two *A* modes $A^1_{1g}$, $A^2_{1g}$, the high-frequency $E^2_g$ mode and a weak additional mode. The latter is presumably a gap mode $A_{gap}$ of Te at sites *VI(1)* and is only observed in the range $1 \leq x \leq 2.3$. Similar phonon modes were observed by Raman scattering from bulk $Bi_2Te_{3-x}Se_x$ crystals.[23] The Raman lines are assigned to the modes from their frequencies and polarization dependence, i.e. the disappearance of modes with $A_g$-symmetry for crossed light polarizations. The vibrational displacement patterns of the atoms for the different modes are sketched in the insets of Figure 6a. Note that the atom at site *VI(2)* is at rest for all modes due to the even symmetry character of the *g*-modes. The mode frequencies are determined by multiple Voigt-curve fits of the Raman spectra and are plotted by symbols in Figure 6b as a function of the Se content in the layers. The nearly constant frequencies of the $A^1_{1g}$ mode at about 62 cm$^{-1}$ and the $E^2_g$ mode at 103 cm$^{-1}$ for $x \leq 1$ qualitatively indicate structural order with Se occupying only the *VI(2)* site. A variation of the mass at site *VI(2)* has no influence on mode frequencies because this site is the resting center of mass for the Raman-active modes with *g*-symmetry.

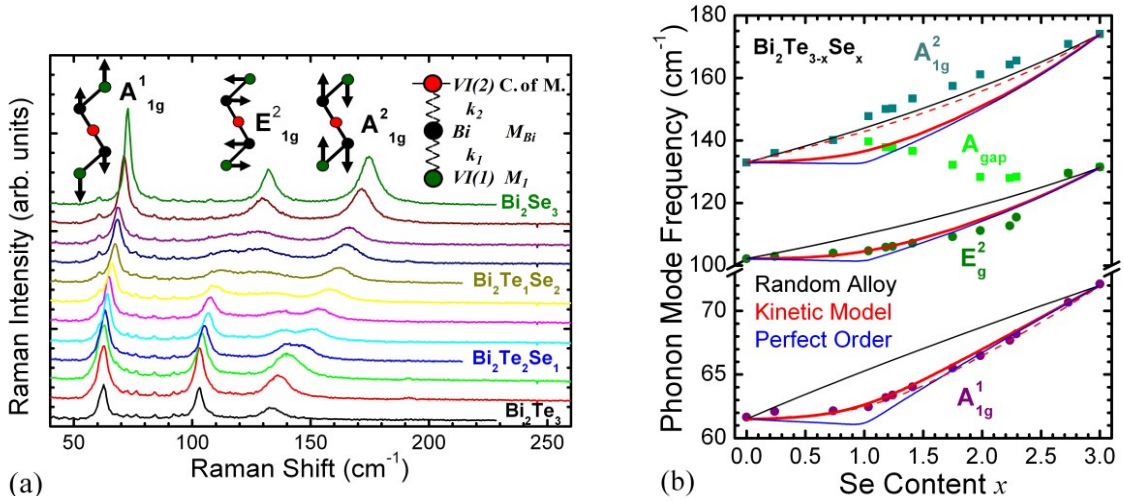

**Figure 6:** (a) Raman spectra taken at room temperature of 70 nm thick Bi$_2$Te$_{3-x}$Se$_x$ layers with varied Se content. The spectra are vertically shifted for clarity. The vibrational displacement patterns of the different phonon modes and a spring-mass model are sketched as insets. (b) Observed phonon mode frequencies depending on Se content and corresponding values calculated with the coupled spring-mass model for random alloying of Se and Te (black curves), perfect order (blue curves) and partial order according to the kinetic model (red curves). Full red curves represent results with spring constants depending only on the Se content of the neighboring sites, while dashed red curves include qualitatively polarization effects due to a redistribution of valence electrons within QLs.

The vibrational coupling between QLs by the very weak van-der-Waals bonds can be neglected. The atomic displacements for Raman-active *g*-modes are symmetric with respect to site *VI(2)* and correspond to those of a *VI(1)-Bi-VI(2)-Bi-VI(1)* molecule with masses and springs dependent on the Se contents of the sites. Mode frequencies are described by a simple model of two springs with constants $k_1$, $k_2$ and masses $M_{Bi}$ and $M_1$ for the Bi atom and site *VI(1)* (see the right-hand inset in Figure 6a. Spring $k_2$ is fixed at site *VI(2)*, the resting center of mass. The equation of motion for *E* or *A* modes reduces to that of two coupled harmonic oscillators given in Equation 3 with analytical solutions for the mode frequencies ω.

$$\begin{pmatrix} M_{Bi}\omega^2 - k_1 - k_2 & k_1 \\ k_1 & M_1\omega^2 - k_1 \end{pmatrix} \cdot \begin{pmatrix} u_{Bi} \\ u_1 \end{pmatrix} = 0 \tag{3}$$

In the virtual crystal approximation, the average mass of site *VI(1)* is given by $M_1 = (1 - x_1)M_{Te} + x_1 M_{Se}$. The spring constants $k_1$ and $k_2$ in $Bi_2Te_3$ and $Bi_2Se_3$ are deduced from the measured mode frequencies in the corresponding binary layers of our series. The spring constants in $Bi_2Te_{3-x}Se_x$ depend on the layer compositions $x_1$ and $x_2$ and are linearly interpolated. The observed frequencies of the $E^2_g$ mode and the $A^1_{1g}$ mode depending on *x* are well described by the spring-mass model with the reasonable assumption that the spring constants $k_i$ depend linearly on the calculated composition of the neighboring layer site *VI(i)*, i.e. $k_1(x_1) = 1.067 + 0.517x_1$ and $k_2(x_2) = 1.672 - 0.036x_2$ (in units of $10^6$ amu·cm$^{-2}$) for *A* modes. The spring constant $k_1$ considerably increases with Se content of site *VI(1)* $x_1$, while $k_2$ is nearly independent of $x_2$. The $E^2_g$ and the $A^1_{1g}$ mode frequencies calculated for $Bi_2Te_{3-x}Se_x$ with partial order as predicted by the kinetic growth model are shown by red curves in Figure 6b. These curves agree well with the experimental results and reproduce the qualitative assumption of nearly constant frequencies for a high degree of order and Se contents $x \leq 1$, as most of the Se atoms with low mass are incorporated at the resting site *VI(2)*. The mode frequencies should even decrease slightly with increasing Se content for $x \leq 1$ in perfectly ordered $Bi_2Te_{3-x}Se_x$ (blue curves). In contrast, monotonic, nearly linear shifts of mode frequencies are expected for a random alloy (black curves). Inspection of the curves shows that both mode frequencies are sensitive to the degree of order, especially around $x = 1$.

Compared to the $A^1_{1g}$ mode, the Raman peaks of the $A^2_{1g}$ mode at $x \approx 1$ and the $E^2_g$ mode at $x \approx 2$ are quite broad and close in frequency to those of the $A_{gap}$ mode, which may reduce the accuracy of the fitted frequency values (Figure 6a). Moreover, these modes may be shifted due to a coupling to the $A_{gap}$ mode with similar frequency. These aspects most probably cannot fully explain the considerable difference between measured $A^2_{1g}$ mode frequencies and those calculated within the simple model with spring constants depending only on the composition of the neighboring site *VI* sublayer. Obviously, the $A^2_{1g}$ mode has a more complex dependence on sublayer compositions and it shows a two-mode behavior at large Se content x ≥ 1. Increased frequencies of the $A^2_{1g}$ mode may be qualitatively explained as follows: The vibrational displacement of Bi layers is in anti-phase to that of the *VI(1)* layers and spring $k_1$ has a predominant influence on the $A^2_{1g}$ mode frequency. The higher electronegativity of Se (2.4) compared to Te (2.1) and Bi (1.8) and the dependence of the sublayer separations on the compositions of both sublayers *VI(1)* and *VI(2)*, suggest that the bond strengths, i. e. the spring constants $k_1$ and $k_2$, also change with the polarity of bonds and thus depend on both sublayer compositions. For $Bi_2Se_3$, the effective charges (Bader charges) of the sites *Se(2)*, *Bi*, and *Se(1)* were calculated by density-functional theory to be -0.83e, +1.0e, and -0.59e, respectively.[24] For $Bi_2Te_3$, corresponding values of effective charges -0.33e, +0.36e, and -0.19e were determined by tight binding calculations.[25] The different effective charges of sites in both materials suggest that the bond polarities in $Bi_2Te_{3-x}Se_x$ will depend on the Se content of the sites. An increase of the Se content $x_2$ of sublayer site *VI(2)* due to ordering causes a partial transfer of valence electrons from neighboring Bi sublayers to this sublayer and the polarity of bonds *Bi-VI(2)* increases. Furthermore, the Bi sublayers receive a larger positive charge and attract electronic charge from the *Bi-VI(1)* bonds, which consequently decrease in polarity.[7] An increase (decrease) of polarity of bonds is expected to decrease (increase) the bond strength and the spring constant, as observed in semiconductors with zinc-blende structure. Consequently, the *A* mode spring constant $k_2$ should decrease and $k_1$ should increase with increasing difference in Se contents between the two sites ($x_2 - x_1$). If we take spring constants $k_1$ and $k_2$ to have additional terms $+0.26(x_2 - x_1)$ and $-0.13(x_2 - x_1)$ (in units of $10^6$ amu·cm$^{-2}$), respectively, the influence of changes in the polarity of the bonds on the *A* mode frequencies is illustrated by the red dashed curves in Figure 6b. The $A^2_{1g}$ mode increases considerably in frequency, while the $A^1_{1g}$ mode behavior is only slightly affected. The agreement with experimental frequencies is improved for both *A* modes. A detailed understanding of $A_g$ phonon modes in (partially) ordered Bi$_2$Te$_{3-x}$Se$_x$, however, requires first-principle calculations of their structural, bonding and vibrational properties.

## 4. Conclusion

We have developed a kinetic growth model to describe the MBE growth of Bi$_2$Te$_{3-x}$Se$_x$ layers under group *VI* rich growth conditions. The model includes adsorption, desorption and incorporation probabilities of Se and Te at the non-equivalent sublayer sites *VI(1)* and *VI(2)* of the quintuple layer *VI(1)-Bi-VI(2)-Bi-VI(1)*. The kinetic rate equations quantitatively predict the Se contents $x_1$ and $x_2$ in the sublayers of sites *VI(1)* and *VI(2)*. They show that despite the highly preferential incorporation of Se at the central site *VI(2)*, the degree of chemical order in Bi$_2$Te$_2$Se$_1$ as quantified by the Se content on this site is kinetically limited to $x_2 = 0.75 \pm 0.04$. The only parameter affecting the chemical order is the site selectivity of Se *s*.

The calculated sublayer compositions were verified by a precise description of the structural properties measured by X-ray diffraction and the phonon mode frequencies measured by Raman spectroscopy for $Bi_2Te_{3-x}Se_x$ layers of all compositions from $x = 0$ to 3. While the lattice parameter $c$ is barely affected by the degree of chemical order, the separation of atomic sublayers depends strongly on the composition of the group *VI* sublayers and thus on the degree of order. The intensity of the partially destructively interfering *0 0 12* X-ray diffraction peak oscillates with increasing Se content due to these non-monotonic variations of sublayer separations and is a sensitive probe of the structural order in $Bi_2Te_2Se_1$ layers.

Our main finding is that the degree of chemical order in MBE-grown $V_2VI_3$ alloys with non-equivalent chalcogen lattice sites is kinetically limited to values well below that expected from the site selectivity of the competing elements, despite this being the most important material-specific parameter for chemical ordering in these material systems. Chemical ordering in other epitaxial crystal structures with non-equivalent lattice sites for competing elements, such as Heusler or oxide compounds, may also be described by this kinetic model with modifications according to the specific material and growth properties.

**Acknowledgements**

We gratefully acknowledge the financial support by the EU ERC-AG Program (project 3-TOP) and the DFG (through SFB 1170 "ToCoTronics"), as well as experimental assistance from N. V. Tarakina. G. K. thanks for the support by the Polish Science Center (Grant No. 2014/14/M/ST3/00484) and by the Foundation for Polish Science by the Master program.

[1]     H. Shi, D. Parker, M.-H. Du, and D. J. Singh, Phys. Rev. Appl. 2015, 3, 014004.

[2]     M. Neupane, S.-Y. Xu, L. A. Wray, A. Petersen, R. Shankar, N. Alidoust, Chang Liu, A. Fedorov, H. Ji, J. M. Allred, Y. S. Hor, T.-R. Chang, H.-T. Jeng, H. Lin, A. Bansil, R. J. Cava, and M. Z. Hasan, Phys. Rev. B 2012, 85, 235406.

[3]     K. Miyamoto, A. Kimura, T. Okuda, H. Miyahara, K. Kuroda, H. Namatame, M. Taniguchi, S.V. Eremeev, T.V. Menshchikova, E.V. Chulkov, K. A. Kokh, and O. E. Tereshchenko, Phys. Rev. Lett 2012, 109, 166802.

[4]     Z. Ren, A. A. Taskin, S. Sasaki, K. Segawa, and Y. Ando, Phys. Rev. B 2011, 84, 165311.

[5]     H. Maaß, S. Schreyeck, S. Schatz, S. Fiedler, C. Seibel, P. Lutz, G. Karczewski, H. Bentmann, C. Gould, K. Brunner, L. W. Molenkamp, and F. Reinert, J. Appl. Phys. 2014, 116, 193708.

[6]     I. Teramoto and S. Takayanagi, J. Phys. Chem. Solids 1961, 19, 124.

[7]     J. R. Drabble and C. H. L. Goodman, J. Phys. Chem. Solids 1958, 5, 142.

[8]     C. E. C. Wood, D. Desimone, K. Singer, and G. W. Wicks, J. Appl. Phys. 1982, *53*, 4230.

[9]     B. Tribollet, A. Benamar, D. Rayane, P. Melinon, and M. Broyer, Z. Phys. D 1993, 26, 352.

[10]    M. T. Litz, K. Watanabe, M. Korn, H. Ress, U. Lunz, W. Ossau, A. Waag, G. Landwehr, Th. Walter, B. Neubauer, D. Gerthsen, and U. Schüssler, J. Cryst. Growth 1996, 159, 54.

[11]    X. W. Zhou, D. K. Ward, J. E. Martin, F. B. van Swol, J. L. Cruz-Campa, and D. Zubia, Phys. Rev. B 2013, 88, 085309.

[12]    J.-L. Mi, M. Brenholm, M. Bianchi, K. Borup, S. Johnsen, M. Sondergaard, D. Guan, R. C. Hatch, P. Hofmann, and B. B. Iverson, Adv. Mat. 2013, 25, 889.

[13]    H. Okuyama, T. Kawasumi, A. Ishibashi, and M. Ikeda, J. Cryst. Growth 1997, 175/176, 587.

[14]     S. Y. Karpov and M. A. Maiorov, Surf. Sci. 1997, 393, 108.

[15]     D. O. Scanlon, P. D. C. King, R. P. Singh, A. de la Torre, S. McKeown Walker, G. Balakrishnan, F. Baumberger, and C. R. A. Catlow, Adv. Mat. 2012, 24, 2154.

[16]     R. W. G. Wyckoff, Crystal Structures Vol. 2, J. Wiley and Sons, New York, USA 1964.

[17]     K. Park, Y. Nomura, R. Arita, A. Llobet, and D. Louca, Phys. Rev. B 2013, 88, 224108.

[18]     G. R. Miller, C.Y. Li, and C. W. Spencer, J. Appl. Phys. 1961, 34, 1398.

[19]     J. R. Wiese and L. Muldawer, J. Phys. Chem. Sol. 1960, 15, 13.

[20]     P. I. Kutnetsov, G. G. Yakushcheva, V. A. Luzanov, A. G. Temiryazev, B. S. Shchamkhalova, V. A. Jitov, and V. E. Sizov, J. Cryst. Growth 2015, 409, 56.

[21]      P. J. Brown, A. G. Fox, E. N. Maslen, M. A. O`Keefe, and B. T. M. Willis, *Internatinal Tables for Crystallography,* Vol. C, 2006 Ch. 6.1. (Springer Netherlands)

[22]     This scaling factor accounts for instrumental geometry factors and structural effects contributing to XRD intensities in thin films.

[23]     W. Richter, H. Köhler, and C. R. Becker, Phys. Stat. Sol. (b) 1977, 84, 619.

[24]     M. S. Christian, S. R. Whittleton, A. Otero-de-la Roza, and E. R. Johnson, Computational and Theoretical Chemistry 2015, 1053, 238.

[25]     P. Pecheur and G. Toussaint, Phys. Lett. A 1989, 135, 223.